\documentclass[10pt]{iopart}
\usepackage{graphicx}  
\def\la{\label}
\def\al{\alpha} 
\def\be{\beta} 
\def\bea{\begin{eqnarray}}
\def\eea{\end{eqnarray}}
\def\bsea{\begin{subeqnarray}}
\def\esea{\end{subeqnarray}} 
\def\beq{\begin{equation}}
\def\eeq{\end{equation}}
\begin{document}

\title[Kinetically driven glassy transition]
{Kinetically driven glassy transition in an exactly solvable 
toy model with reversible mode coupling mechanism and trivial 
statics}

\author{Bongsoo Kim\dag\ and Kyozi Kawasaki\ddag  
\footnote[3]{Permanent address: 4-37-9 Takamidai, Higashi-ku, Fukuoka
  811-0215, Japan}} 

\address{\dag\ Department of Physics, Changwon National University, 
Changwon, 641-773, Korea}

\address{\ddag\ CNLS, Los Alamos National Laboratory, 
Los Alamos, NM 87545, USA}

\begin{abstract}
We propose a toy model with reversible mode coupling mechanism and with
trivial Hamiltonian (and hence trivial statics).
The model can be analyzed exactly without relying upon uncontrolled
approximation such as the factorization approximation employed in the current MCT.
We show that the model exhibits a kinetically driven transition from an ergodic phase to 
nonergodic phase. The nonergodic state is the  nonequilibrium stationary solution of
 the Fokker-Planck equation for the distribution function of the model.

\end{abstract}




\section{Introduction}

First-principles understanding on the rich dynamic phenomena and the nature of the 
liquid-glass transition still remains as a challenging aim \cite{ean}.
As the only existing first-principle theory, the mode coupling theory (MCT) of 
supercooled liquids and the glass transition enjoyed considerable success in
describing the dynamics of weakly supercooled regime of liquids \cite{mct}.
Notwithstanding this surprising success, there are the following several 
unresolved issues concerning the basis of MCT:
(a) A crucial ingredient of MCT is the factorization approximation which replaces
the four-body time correlation functions by the product of two-body time correlation
functions. This approximation is completely uncontrolled and its region of  validity is 
{\it a priori} unknown. 
(b) The idealized MCT predicts a sharp dynamic transition to a nonergodic state at
a certain temperature. But MCT does not provide any information on the nature of this 
nonergodic state.
(c) The physical picture of the so called hopping processes in an extended version 
of MCT is still lacking.

In recent years, possible deep connection between the structural glass
and a class of spin glass models has been pointed out \cite{kirk}. 
In particular, the Langevin dynamics of the spherical $p$-spin model 
can be analyzed exactly  in the thermodynamic limit due to the mean field nature of the model
(i.e., full connectivity of the spins) \cite{cris,ck}. 
This analysis shows that the dynamic equation for the spin correlation 
function in equilibrium for $p=3$ has the same form as 
in the Leutheusser's schematic mode coupling equation for the density correlator \cite{leuth}.
The sharp dynamic transition observed in this class of models are
driven by  the dissipative nonlinearity in the equation of motion which
originates from the nonlinear Hamiltonian  \cite{trieste}. 
In contrast to this, the glassy behavior in the above-mentioned MCT  (as well as in our model 
given below) is driven by the {\em reversible} nonlinearity \cite{kk} which is dynamically
generated   and hence a non-trivial Hamiltonian is not necessary.  

With these situations, we thought that it is important to develop
a toy model with following three ingredients:
\begin{itemize}
\item reversible mode coupling mechanism
\item trivial statics
\item mean-field type so that the model can be exactly solvable.
\end{itemize}
We have proposed such a toy model in a recent publication \cite{kkbk}. 
Here we further analyze the model.
The model yields the self-consistent equations for the relevant 
correlation functions of the type familiar in the mode coupling theories of
supercooled liquid and glass transition, where the strength of the hopping
processes can be readily tuned.
In the sense that the glassy behavior in this toy model
is induced  by the kinetics of the reversible mode coupling mechanism,
our model is similar in spirit  to the kinetically constrained models, the theme of 
the present workshop.

\section{Model}

Our model is defined as the following Langevin equations for the 
$N$-component density variable $a_i(t)$ with $i=1,2,\cdots,N$ and 
the $M$-component velocity variable $b_{\al}$ with $\al=i,2, \cdots, M$.
Here and after we will use Roman indices for the component of $a$ and 
Greek for that of $b$.
\bea
\dot{a}_i = K_{i\al}b_{\al}+\frac{\omega}{\sqrt{N}} J_{ij\al}a_j b_{\al} \la{EQ1} \\ 
\dot{b}_{\al}  = -\gamma b_{\al}-\omega^2 K_{j\al}a_j   
-\frac{\omega}{\sqrt{N}}J_{ij\al} (\omega^2a_ia_j- T\delta_{ij})+f_{\al}  \la{EQ2} \\
 <f_{\al}(t)>=0, \qquad <f_{\al}(t)f_{\be}(t')> = 2\gamma T\delta_{\al \be} \delta (t-t') \la{EQ3}
\eea
where the summation is implied for repeated indices.
Here $\gamma$ is the decay rate of the velocity $b_{\al}$ and $\omega$ gives the $j$-independent 
frequency of oscillation of the density $a_j$. 
The thermal noise $f_{\al}(t)$  are independent gaussian random variables with zero mean
 and variance $2\gamma T$, $T$ being the temperature of the heat bath 
with which the system has a thermal contact.
The choice of this variance guarantees the proper equilibration of the system.
The $N \times M$ matrix $K_{i\al}$ plays an important role in the model and 
for later purpose we impose the (one-sided) orthogonality
\beq
K_{i\al}K_{i\be}=\delta_{\al\be}, \qquad K_{i\al}K_{j\al}\neq \delta_{ij}
\la{EQ4}
\eeq
where the last equation is due to the inequality $M<N$.
For $M=N$ we can impose an additional condition $K_{i\al}=\delta_{i\al}$ and hence trivially
 $K_{i\al}K_{j\al}=\delta_{ij}$.
We also note that  $K_{i\al}$ governs linearized reversible dynamics of the model 
with the dynamical matrix $\bf \Omega$ given by $\Omega_{ij} \equiv \omega^2K_{i\al}K_{j\al}$. 
The reversible nonlinear mode coupling terms are the ones involving 
the mode coupling coefficients $J_{ij\al}$ which are chosen to be quenched (time-independent) gaussian 
random variables with the following properties:
\bea
\overline{J_{ij\alpha}}^J &=0,  \nonumber \\
\overline{J_{ij\alpha}J_{kl\beta}}^J &= \frac{g^2}{N}
\biggl[(\delta_{ik}\delta_{jl}+\delta_{il}\delta_{jk})\delta_{\alpha
\beta} + K_{i\beta}(K_{k\al}\delta_{jl}+K_{l\al}\delta_{jk}) \nonumber \\  
&+ K_{j\beta}(K_{k\al}\delta_{il}+K_{l\al}\delta_{ik})\biggr] \la{EQ5}
\eea
where $\overline{\cdots }^J$ denotes average over the $J$'s.
Note that there is no thermal noise which acts directly on the density variable in (\ref{EQ1}).
This is because the model is constructed so as to mimic the dynamics of fluid. 
Equation (\ref{EQ1}) is analogous to the equation of continuity of fluid and 
Eq.(\ref{EQ2}) is like the equation of motion where the right hand side is like
the force acting on a fluid element.
In constructing this model, we were motivated by the works \cite{kraich,bouch} in which
random coupling models involving an infinite component order parameter have been shown 
to be exactly analyzed by mean-field-type concepts.
We will thus eventually take $N$ and $M$ infinite with the ratio
$\delta^* \equiv M/N$ kept finite.

One can derive from the Langevin equations (1)-(3) the corresponding Fokker-Planck equation 
for the probability distribution function $D(\{a\},\{b\},t)$ for
our variable set denoted as  $\{a\},\{b\}$  as follows
\beq
\partial_t D(\{a\},\{b\},t) =  {\hat L} D(\{a\},\{b\},t) \la{EQ6}
\eeq
where the Fokker-Planck operator is given by 
${\hat L}={\hat L}_0 + {\hat L}_1+ {\hat L}_{MC}$ \\
with  
\beq
\eqalign{{\hat L}_0 \equiv  \frac{\partial}{\partial b_{\alpha}} 
\gamma \left(T \frac{\partial}{\partial
b_{\alpha}}+b_{\alpha} \right), \quad
 {\hat L}_1 \equiv K_{j \al} \left(- \frac{\partial}{\partial a_j}
b_{\al}+\frac{\partial}{\partial b_{\al}} \omega^2
a_j \right), \nonumber \\
 {\hat L}_{MC}  \equiv \frac{1}{\sqrt{N}} J_{ij\alpha}\left(
-\frac{\partial}{\partial a_i}\omega a_jb_{\alpha} +
\frac{\partial}{\partial b_{\alpha}} \omega (\omega^2
a_ia_j-T\delta_{ij})\right)} \la{EQ7}
\eeq
It is then easy to show that the {\it equilibrium} stationary distribution 
(i.e., ${\hat L}D_e({a},{b})=0$) is given by
\beq
 { D}_e(\{a\},\{b\}) = cst. \rme^{-\sum_{j=1}^N\frac{\omega^2}{2T}a_j^2-
\sum_{\alpha=1}^M\frac{1}{2T}b_{\alpha}^2} \la{EQ8}
\eeq
where $cst.$ is the normalization factor.

\section{Analysis and discussion}

For the subsequent analysis it is most convenient to introduce 
 the following generating functional 
\beq
\fl \hat Z\{h^a,{\hat h}^a,h^b,{\hat h}^b \} \equiv \int d\{a\}\int
d\{b\}\int d\{\hat a\}\int d\{\hat b\} \rme^{\rmi\int dt (h_j^aa_j+\hat
h_j^a\hat a_j +h_{\alpha}^bb_{\alpha}+ \hat h_{\alpha}^b\hat
b_{\alpha})} \rme^{\hat{\cal S}} \la{EQ9}
\eeq
where the integrals are the functional integrals over the variable sets
 $\{a\},\{\hat a \},\{b\},\{\hat b \}$ and
the $h$'s and the $\hat h$'s the conjugate source fields.
The action $\hat{\cal S}$ was decomposed into two parts $\hat{\cal S}_0$ and $\hat{\cal S}_I$ 
which take the form
\bea
\fl \hat{\cal S}_0 = \int dt\left\{ \rmi \hat
a_i\big(\dot{a}_i-K_{i\alpha} b_{\alpha}\big) +\rmi \hat
b_{\alpha}\big(\dot{b}_{\alpha} +\gamma b_{\alpha}+\omega^2K_{i\alpha}
a_i-f_{\alpha}\big)\right\}(t) \la{EQ10} \\ 
\hat{\cal S}_I=J_{jk\alpha}\hat X_{jk\alpha} \la{EQ11} \\
\hat X_{jk\alpha} \equiv   \frac{\omega}{\sqrt{N}} \int dt \left
\{-\rmi \hat a_j a_k b_{\alpha} + \rmi \hat b_{\alpha}\omega^2
a_j a_k\right\}(t) \la{EQ12}
\eea
where we have dropped the term $T\delta_{ij}$ coming from  eq.(\ref{EQ2})
since this term is negligible in the limit of infinite $M$ and $N$. 
The functional determinant associated with the Langevin equations (\ref{EQ1})-(\ref{EQ3}) which
should appear in the integrand of the generating functional $\hat Z$ 
was equated to unity assuming the It\^{o} calculus \cite{fh}. 
The various correlation functions and response functions are obtained by
 taking various functional derivatives of $\ln Z\{h^a,{\hat h}^a,h^b,{\hat h}^b \}$
 with respect to $h$'s and  $\hat h$'s and setting them equal to zero
 in the end in the standard way, where $Z$ is the generating functional $\hat Z$, 
 averaged over the $f$'s and the $J$'s.

We now note that the replacements $ \rmi \hat a_j \rightarrow
(\omega^2/T)a_j, \quad \hat b_{\alpha} \rightarrow b_{\alpha}/T$
in  $\hat X_{jk\al}$ leads to $\hat X_{jk\al}=0$. 
Hence we can rewrite $\hat X_{jk\al}$ also as
\beq
 \hat X_{jk\al}=\tilde X_{jk\al}\equiv  \frac{\omega}{\sqrt{N}} \int dt
\left\{-\rmi \tilde a_j a_k b_{\alpha}+\rmi \omega^2 \tilde b_{\alpha}a_ja_k\right\}(t) \la{EQ13}
\eeq
where $\rmi \tilde a_i \equiv \rmi \hat a_i +(\omega^2/T)a_i$ and 
$\rmi b_{\al} \equiv \rmi \hat b_{\al}+ b_{\al}/T$.

We now obtain  for this toy model the equilibrium correlation functions defined as
\beq
\eqalign{\fl C_a(t-t') \equiv  \frac{1}{N} <a_j(t)a_j(t')>,  \quad
 C_{ab}(t-t') \equiv \frac{1}{M}K_{j\al}<a_j(t)b_{\al}(t')>,  \\
 \fl C_{ba}(t-t') \equiv \frac{1}{M}K_{j \al}<b_{\al}(t)a_j(t')>,  \quad
  C_b(t-t') \equiv \frac{1}{M} <b_{\alpha}(t)b_{\alpha}(t')>,  \\
  C_a^K(t-t') \equiv \frac{1}{M}K_{i\al}K_{j\al}<a_i(t)a_j(t')>} \la{EQ14}
\eeq
It turns out that we need to have the last correlation function to close the self-consistent
 set of equations for the correlators when $M<N$.
Note that for the case $M=N$, if $K_{i\al}=\delta_{i\al}$ is imposed, then $C_a^K(t-t')=C_a(t-t')$.
The corresponding response functions can be defined as 
\beq
\eqalign{\fl G_a(t-t') \equiv  \frac{1}{N} <a_j(t)\rmi \hat a_j(t')>,  \quad
 G_{ab}(t-t') \equiv \frac{1}{M}K_{j\al}<a_j(t)\rmi \hat b_{\al}(t')>,  \\
\fl G_{ba}(t-t') \equiv \frac{1}{M}K_{j \al}<b_{\al}(t) \rmi \hat a_j(t')>,  \quad
  G_b(t-t') \equiv \frac{1}{M} <b_{\alpha}(t)\rmi \hat b_{\alpha}(t')>,  \\
  G_a^K(t-t') \equiv \frac{1}{M}K_{i\al}K_{j\al}<a_i(t)\rmi \hat a_j(t')>} \la{EQ15}  
\eeq
Since we have a Gaussian stationary solution, we get the fluctuation-dissipation relationships 
(FDR) of the form \cite{haake}
\beq
\eqalign{\fl G_a(t-t')=-\theta(t-t')\frac{\omega^2}{T}C_a(t-t'), \quad
G_{ab}(t-t')=-\theta(t-t')\frac{1}{T}C_{ab}(t-t'), \\
\fl G_{ba}(t-t')=-\theta(t-t')\frac{\omega^2}{T}C_{ba}(t-t'), \quad
G_b(t-t')=-\theta(t-t')\frac{1}{T}C_b(t-t'), \\
G_a^K(t-t')=-\theta(t-t')\frac{\omega^2}{T}C_a^K(t-t')} \la{EQ16}
\eeq
where $\theta(t)$ is the unit step function: $\theta(t)=1$ for $t\geq 0$ and $0$ otherwise.
Note that this form of the FDR is rather unusual since the FDR usually takes the form
$G(t)=-\theta(t)\partial_t C(t)/T$.

Another useful property arising from the causality and the above FDR is 
the following property 
\beq
<\hat A(t) X(t') > = < X(t) \tilde A (t') > =0 \quad \mbox{for}  \quad t \geq t' \la{EQ17}
\eeq
for $A(t)=(a(t), b(t))$ and an arbitrary function
$X(t)=X(a(t),b(t), \hat a(t), \hat b(t))$. 

We now take averages of $\hat Z$ over the thermal noise $f_{\al}$
and the quenched random coupling $J_{ij\al}$. In so doing
we use the following properties which hold for the gaussian random variables:
\beq
\eqalign{\left<\rme^{-\rmi \int dt \hat b_{\al}(t) f_{\al}(t)}\right>
=\rme^{-\gamma T \int dt \hat b_{\al}(t)^2} \nonumber \\
\overline{\rme^{ J_{jk\al}\hat X_{jk\al}}}^J =\rme^{\frac{1}{2}
\overline{ J_{jk\al} J_{lm\be} }^J \hat X_{jk\al}\hat X_{lm\be}}} \la{EQ18}
\eeq
Defining the actions $S_0$ and $S_I$ as
\beq
\rme^{{\cal S}_0} \equiv \left< \rme^{\hat S_0} \right>, \qquad
\rme^{{\cal S}_I} \equiv \overline{\rme^{\hat S_I}}^J,  \la{EQ19}
\eeq
we obtain 
\bea
\fl {\cal S}_0 = \int dt \left\{ \rmi\hat a_i (\dot a_i -K_{i\al}b_{\al})(t)
+\rmi \hat b_{\al} (\dot b_{\al}+\gamma b_{\al}+
\omega^2 K_{i\al}a_i )(t) - \gamma T \hat b_{\al}^2(t) \right\} \nonumber \\
\fl = \int dt \left\{ \rmi\hat a_i\biggl(\frac{T}{\omega^2}
\rmi\dot{\tilde a}_i-T K_{i\alpha}\rmi\tilde b_{\alpha}\biggr)(t) +\rmi\hat
b_{\alpha} \biggl(T\rmi\dot{\tilde b}_{\alpha}+ T K^T_{\alpha i}
\rmi{\tilde a}_i +\gamma T \rmi\hat b_{\alpha} \biggr)(t) \right\} \la{EQ20}  
\eea
where the last line is obtained using the property (\ref{EQ17}).
Now we have to deal with the interaction part
$S_I=\overline{J_{jk\al}J_{lm\be}}^J{\hat X}_{jk\al}{\hat X}_{lm\be}/2$.
One can show that in the limit of $M,N \rightarrow \infty$ 
fluctuations can be neglected so that quantities like
$a_j(t)a_j(t')/N$ etc. are replaced by $C_a(t,t')$, etc.
The interaction part ${\cal S}_I$ then becomes gaussianized in the limit of 
$M,N \rightarrow \infty$.
The final expression for  ${\cal S}_I$ is then given by
\bea
{\cal S}_I &= \int dt \biggl\{i\hat a_i(t) \frac{T}{\omega^2}
\Sigma_{aa} \otimes i\tilde a_i(t)+ K_{i\al} i\hat a_i(t) T
 \Sigma_{ab} \otimes i\tilde b_{\al}(t) \nonumber \\
 &+ K_{i\al}i\hat b_{\alpha}(t) \frac{T}{\omega^2}\Sigma_{ba}
\otimes i\tilde a_i(t) +i\hat b_{\alpha}(t)T\Sigma_{bb}
 \otimes i\tilde b_{\alpha}(t) \biggr\} \la{EQ21}
\eea 
where $\Sigma \otimes a(t) \equiv \int_{-\infty}^t dt' \,
\Sigma(t-t')a(t')$ etc.
Here the kernels $\Sigma$'s are given by
\beq
\eqalign{\fl \Sigma_{aa}(t-t') \equiv \delta^*\frac{g^2\omega^4}{T}\bigl(C_a(t-t')C_b(t-t')+
\delta^* C_{ab}(t-t')C_{ba}(t-t')\bigr), \\
\Sigma_{ab}(t-t') \equiv -2 \delta^* \frac{g^2\omega^4}{T}C_a(t-t')C_{ba}(t-t')  \\
\Sigma_{ba}(t-t') \equiv -2 \delta^* \frac{g^2\omega^6}{T}C_a(t-t')C_{ab}(t-t'),  \\
 \Sigma_{bb}(t-t')\equiv \frac{2g^2\omega^6}{T}C_{a}(t-t')^2} \la{EQ22}
\eeq  
Three kernels $\Sigma_{aa}$, $\Sigma_{ab}$, and $\Sigma_{ba}$ comes from the 
nonlinear coupling term in the original Langevin equation (\ref{EQ1}), and 
the kernel $\Sigma_{bb}$ arises from the density nonlinearity in (\ref{EQ2}).
We also note that the correlator $C_a^K(t,t')$ is not involved in the $\Sigma$'s.

From the effective gaussian action ${\cal S}_{eff} \equiv {\cal S}_0 + {\cal S}_I$
we can readily write down the following linearized Langevin equations 
for $a_i$ and $b_{\al}$
\bea
\fl \dot{a}_i(t)&= K_{i\alpha} b_{\al}(t)- \Sigma_{aa} \otimes a_i(t)
-K_{i\al} \Sigma_{ab}\otimes b_{\al}(t)+f^a_i(t)  \la{EQ23} \\
\fl \dot{b}_{\al}(t)&=-\gamma b_{\al}(t)-\omega^2
 K_{i\al}a_i(t) -K_{i\al} \Sigma_{ba}\otimes a_i(t) -
 \Sigma_{bb} \otimes b_{\alpha}(t)+ f^b_{\alpha}(t) \la{EQ24}
\eea
where $f^a$ and $f^b$ are the effective thermal noises whose correlations
are given by
\beq
\eqalign{<f^a_i(t)f^a_j(t')>&= \frac{T}{\omega^2}
[\Sigma_{aa}(tt')+\Sigma_{aa}(t't)]\delta_{ij} \\
 <f^a_i(t)f^b_{\al}(t')>&= K_{i\al} T[\Sigma_{ab}(tt')+
 \frac{1}{\omega^2}\Sigma_{ba}(t't)] \\
 <f^b_{\al}(t)f^a_i(t')>&= K_{i\al} T[\Sigma_{ba}(tt')+
  \frac{1}{\omega^2} \Sigma_{ab}(tt')]  \\
 <f^b_{\alpha}(t)f^b_{\beta}(t')>&= \biggl( 2\gamma T\delta (t-t')
 +T[\Sigma_{bb}(tt')+\Sigma_{bb}(t't)] \biggr) \delta_{\al \be}} \la{EQ25}
\eeq

Now we are ready to obtain a set of self-consistent equations for the {\em five} correlators 
from the linearized Langevin equations.
By multiplying (\ref{EQ23}) by $a_i(0)/N$ and (\ref{EQ24}) by $K_{i\al}a_i(0)/M$ and averaging over the
effective thermal noise, we obtain 
\bea
\dot C_a(t) &= \delta^*C_{ba}(t)
-\Sigma_{aa}\otimes C_a(t)-\delta^*\Sigma_{ab}\otimes C_{ba}(t) \la{EQ26} \\
\dot C_{ba}(t) &= -\gamma C_{ba}(t)-\omega^2 C_a^K(t)
-\Sigma_{ba}\otimes C_a^K(t)-\Sigma_{bb}\otimes C_{ba}(t) \la{EQ27}
\eea
where we used the causality requirements
$<f^a_i(t)a_i(0)>=0$ and $K_{i\al}<f^b_{\al}(t)a_i(0)>=0$.
Note that the correlator $C_a^K(t)$ appears in the equation for $C_{ba}(t)$.
In order to obtain the equation for $C_a^K(t)$, we multiply (\ref{EQ23}) by
$K_{i\be}K_{j\be}a_j(0)/N$ and take thermal average. Then we obtain
\beq
\dot C_a^K(t) = C_{ba}(t)
-\Sigma_{aa}\otimes C_a^K(t)-\Sigma_{ab}\otimes C_{ba}(t) \la{EQ28}
\eeq
Similarly by multiplying (\ref{EQ23}) and (\ref{EQ24}) by $K_{i\be}b_{\be}(0)/M$ and 
$b_{\al}(0)/M$, respectively, and performing the thermal average
we obtain the following equations for $C_{ab}(t)$ and $C_b(t)$
\bea
\dot C_{ab}(t) &= C_b(t)
-\Sigma_{aa}\otimes C_{ab}(t)-\Sigma_{ab}\otimes C_b(t), \la{EQ29} \\
\dot C_b(t) &= -\gamma C_b(t)-\omega^2 C_{ab}(t)
-\Sigma_{ba}\otimes C_{ab}(t)-\Sigma_{bb}\otimes C_b(t) \la{EQ30}
\eea
The equations (\ref{EQ26})-(\ref{EQ30}) constitute the self-consistent equations for the 5 correlators
$C_a(t)$, $C_{ba}(t)$, $C_a^K(t)$, $C_{ab}(t)$, and $C_b(t)$.
This set of equations can be solved numerically with the initial conditions
$C_a(0)=C_a^K(0)=T/\omega^2$, $C_{ab}(0)=C_{ba}(0)=0$, and $C_b(0)=T$.

In analytic side, it is very convenient to work with the equations of the Laplace transformed
correlation functions defined as
$C^L(z) \equiv \int_0^{\infty} dt\, \rme^{-zt} \, C(t)$.
Performing the Laplace transformation of the self-consistent equations we obtain
\bea
 zC^L_a(z)&= \frac{T}{\omega^2} + (1-\Sigma^L_{ab}(z))\delta^* C^L_{ba}(z) 
- \Sigma^L_{aa}(z)C^L_a(z)   \la{EQ31} \\
 zC^L_{ba}(z)&=-(\gamma + \Sigma^L_{bb}(z)) C^L_{ba}(z) 
 -( \omega^2 +\Sigma^L_{ba}(z)) C^{KL}_a(z) \la{EQ32}   \\
 zC_a^{KL}(z)&=\frac{T}{\omega^2}+(1-\Sigma^L_{ab}(z)) C^L_{ba}(z) 
-\Sigma_{aa}^L(z)C_a^{KL}(z) \la{EQ33} \\
 zC^L_{ab}(z)&= (1-\Sigma^L_{ab}(z)) C^L_b(z) 
- \Sigma^L_{aa}(z) C^L_{ab}(z) \la{EQ34} \\
 zC^L_b(z)&=T - (\omega^2 +\Sigma^L_{ba}(z))C^L_{ab}(z) 
-(\gamma + \Sigma^L_{bb}(z)) C^L_b(z)\la{EQ35}
\eea
From (\ref{EQ31})-(\ref{EQ33}), we obtain $C^L_a(z)$, $C^{KL}_a(z)$, and $C^L_{ba}(z)$
in terms of $\Sigma$'s as follows:
\bea
\fl C^L_a(z)=\frac{T}{\omega^2}\frac{1}{z+\Sigma^L_{aa}(z)}
\biggl[1-\delta^* \frac{\omega^2(1-\Sigma^L_{ab}(z))^2}{ 
(z+\Sigma^L_{aa}(z))(z+\gamma+\Sigma_{bb}(z))+ \omega^2(1-\Sigma^L_{ab}(z))^2} \biggr] \la{EQ36} \\
 C^{KL}_a(z)=\frac{T}{\omega^2}\biggl[z+\Sigma^L_{aa}(z)
+\frac{\omega^2(1-\Sigma^L_{ab}(z))^2} {z+\gamma+\Sigma_{bb}(z)} \biggr]^{-1} \la{EQ37} \\
C^L_{ba}(z)=-\frac{T(1-\Sigma^L_{ab}(z))}{(z+\Sigma^L_{aa}(z))
(z+\gamma+\Sigma_{bb}(z)) +\omega^2(1-\Sigma^L_{ab}(z))^2} \la{EQ38}
\eea
Here we  have used the following symmetry relation
\beq
\Sigma^L_{ba}(z)=-\omega^2\Sigma^L_{ab}(z) \la{EQ39}
\eeq
which follows from the definition of the kernels $\Sigma_{ab}$ and $\Sigma_{ba}$, (\ref{EQ22}), and 
$C_{ab}(t)\equiv K_{i\al}<a_i(t)b_{\al}(0)>=K_{i\al}<a_i(0)b_{\al}(-t)>
=-K_{i\al}<b_{\al}(t)a_i(0)>=-C_{ba}(t)$.
The first equality is due to the time translation invariance and the second one
from the time reversal property of the velocity components.
Note that for $\delta^*=1$ the two correlators $C^L_a(z)$ and $C^{KL}_a(z)$ become identical.

Similarly, from (\ref{EQ34})-(\ref{EQ35}), we obtain
\bea
C^L_{ab}(z)&=\frac{T(1-\Sigma^L_{ab}(z))}{(z+\Sigma^L_{aa}(z))
(z+\gamma+\Sigma_{bb}(z)) +\omega^2(1-\Sigma^L_{ab}(z))^2}\la{EQ40} \\
C^L_b(z)&=\frac{T(z+\Sigma^L_{aa}(z))}{(z+\Sigma^L_{aa}(z))
(z+\gamma+\Sigma_{bb}(z)) +\omega^2(1-\Sigma^L_{ab}(z))^2}\la{EQ41}
\eea

Now let us look at the behavior of the correlators for different values of $\delta^*$.
For $\delta^*=0$ the only nonvanishing kernel is $\Sigma^L_{bb}(z)$.
Hence we obtain
\bea
C^L_a(z)&=\frac{T}{\omega^2}\frac{1}{z}, \quad \Sigma^L_{bb}(z)=\frac{2g^2\omega^2T}{z}, \la{EQ42} \\
C^{KL}_a(z)&=\frac{T}{\omega^2}\frac{1}{z}\left[1-\frac{\omega^2}
{z(z+\gamma)+(1+2g^2T)\omega^2}\right], \la{EQ43} \\
C^L_b(z)&=\frac{zT}{z(z+\gamma)+(1+2g^2T)\omega^2}, \la{EQ44} \\ 
C^L_{ab}(z)&=-C^L_{ba}(z)=\frac{T}{z(z+\gamma)+(1+2g^2T)\omega^2}\la{EQ45} 
\eea
Here we point out that there appears to be a subtlety associated with the two 
limiting procesures: (A) first take $\delta^*=M/N=0$ before any calculation.
(B) first calculate with $\delta^*>0$ and then take the limit $\delta^* \rightarrow 0+$.
The procesure (A) gives both $C_a(t)=C_a(0)=T/\omega^2$ and 
$C^K_a(t)=C^K_a(0)=T/\omega^2$. This is simply due to the fact that 
the $\{a\}$ variables are time-independent since
there is no velocity variable $\{b\}$ that drives dynamics of $\{a\}$.
However the results (\ref{EQ42})-(\ref{EQ45}) were obtained by adopting the second limiting 
procedure (B). Here $C_a(t)$ is trivially nonergodic: $C_a(t)=C_a(0)=T/\omega^2$ whereas
and $C^K_a(t)$ exhibits a nontrivial nonergodic behavior: 
$C^K_a(t\rightarrow \infty)=(T/\omega^2)\cdot 2g^2T/(1+2g^2T)$.
The difference between these two procedures can be seen also by looking at (\ref{EQ33})
for $C^{KL}_a(z)$. The terms except the first one on the right hand side is absent if the 
first limiting procedure (A) is adopted, whereas it remains finite in the second
limiting procedure (B).

For $\delta^*=1$ where $M=N$ and $K_{i\al}=\delta_{i\al}$, 
$C^L_a(z)=C^{KL}_a(z)$ reproduces the equation derived in \cite{sdd}, apart from
the wave number dependence.
Note that if we put $\Sigma^L_{aa}(z)=\Sigma^L_{ab}(z)=0$ {\it by hand},
(\ref{EQ36}) or (\ref{EQ37}) gives a closed equation for $C_a(t)$ alone. This equation is nothing but
the Leutheusser's schematic MC equation giving a dynamic transition from
ergodic phase to nonergodic one. But in reality $\Sigma_{aa}$ and 
$\Sigma_{ab}$ can not be ignored and our numerical solution strongly indicates
that  the system remains ergodic for all temperatures due to the 
strong contribution of these so called hopping terms. 
Furthermore these hopping terms do not become self-consistently small 
as temperature is lowered. Therefore the 
density correlator does not show a continuous slowing down with lowering temperature.
This result was   striking to us  since usually a mean-field-type theory, such as
the dynamics of the spherical $p$-spin model in the limit of $N \rightarrow \infty$, 
often gives a sharp dynamic transition. 
In fact, we were first constructing the toy model with $M=N$ and we expected that 
the model designed to rigorously reproduce the idealized MCT exhibits such a dynamic
transition. But to our surprise the dynamic transition was absent in the
$N$-component toy model. 
This aspect is  a fundamental difference in the two kinds of mean-field-type
theories with and without reversible mode coupling. The foremost example of the latter
is the sherical $p$-spin model where the ergodic-to-nonergodic transition is driven by
the dissipative nonlinearity which comes from the nonlinear random Hamiltonian.
As demonstrated below, in order  to have such a sharp transition in our toy model, 
 we find it necessary  to extend the original $N$-component model to the model with $M<N$.
Thus it is very difficult to understand the idealized MCT {\em without} relying upon
uncontrolled approximation. 
It is also interesting to note that the ergodicity restoring process
in our toy model (represented by the kernels $\Sigma_{aa}$ and $\Sigma_{ab}$) 
has nothing to do with a thermally activated energy barrier crossing since the gaussian 
Hamiltonian in our model does not possess such a barrier.

\begin{figure}
\begin{center}
\includegraphics*[width=12.5cm,height=9cm]{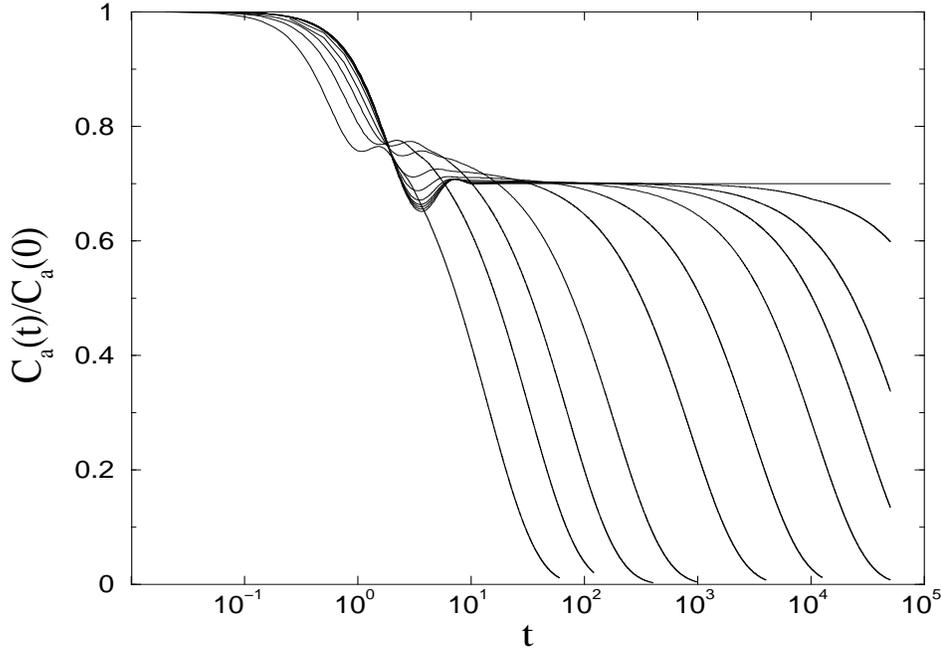}
\end{center}
\caption{\label{fig1} The relaxation of the  nonmalized density correlator $C_a(t)/C_a(0)$ for 
$\delta^*=0.3$. The other parameters are given by $g=\gamma=\omega=1$.
The curves are, from left to right at long times, for $T=5,2,1,0.5,0.2,0.1,0.05,0.02,0.01$, 
and $0.001$}
\end{figure}

Our numerical solution for $\delta^*=0.3$ is shown in Figure 1 with various values of 
$T$. The other parameters were fixed as $\omega=1$, $\gamma=1$, and $g=1$.
As $T$ is lowered, the relaxation exhibits a continuous slowing and it appears to be frozen 
at lowest $T$.  One may ask whether this  freezing reflects the presence of the genuine nonergodicity or
it is merely apparent: the decaying will be observed if the observation time window is further extended.
The question of the existence of nonergodicity is easily answered 
in the usual idealized MCT where one can easily solve the closed equation for the nonergodicity
parameter to obtain the phase diagram.
The situation is very different in our toy model.
When we expand the correlators as $C^L_a (z)=f_a/z+f^{(0)}_a + f^{(1)}z +\cdots$ etc.,
we end up with a herarchically connected set of equations for all the $f$'s, which can not
be easily analyzed numerically.

An analytic feature signifying the presence of the genuine nonergodic state can be seen  by 
adiabatically eliminating the velocity components in the
limit of large $\gamma$ and obtaining the Fokker-Planck equation
for the distribution function $\tilde D (\{a\},t)$
containing only the $\{a\}$ variables:
\beq
\frac{\partial \tilde D(\{a\},t)}{\partial t} =
 \frac{\partial}{\partial a_i}
\left[ Q_{ij}(\{a\}) \left( \frac{\partial }{\partial a_j}
+\frac{\omega^2}{T}a_j \right) \tilde D(\{a\},t) \right] \la{EQ47}
\eeq
Here the diffusion matrix $Q_{ij}(\{a\})$ is given by
\bea
Q_{ij}(\{a\})\equiv \frac{T}{\gamma} M_{i\al} M_{j\al} \nonumber \\
M_{i\al} \equiv K_{i\al}+ \frac{\omega}{\sqrt{N}}J_{ik\al}a_k \la{EQ48}
\eea
An important point is that the diffusion matrix $Q_{ij}$ is {\em
singular} for $M<N$, i.e., det$|{\bf Q}|=0$ \cite{risken}. 
The proof is simple. Define a $N \times N$ matrix
${\bf M}$ by ${\bf M}_{ij} \equiv M_{i,j=\al}$ for $j \leq M$, 
${\bf M}_{ij} \equiv 0$ for $j > M$. Then we obtain in matrix notation
${\bf Q}=(T/\gamma) {\bf M}\cdot{\bf M}^T$ (The superscript $T$ denotes the
transposed matrix). Then det$|Q|=(T/\gamma)^N$(det$|{\bf M}|)^2 =0$
since det$|{\bf M}|=0$ by construction.
This implies that the Fokker-Planck equation (\ref{EQ47})
can have {\em nonequilibrium} stationary solution other than 
the equilibrium one, $\tilde D_e(\{a\})=cst. \exp (-\omega^2 a_j^2/2T)$.
This nonequilibrium stationary solutions are precisely the kind of 
nonergodic states found numerically in the present toy model.
The general stationary solution \cite{merida} is given by
\beq
\tilde D_L(\{a\})={\cal F}(\xi_j a_j)\,\rme^{-\frac{\omega^2}{2T}a_i^2} \la{EQ49}
\eeq
where $\xi_i$ is the eigenvector of the diffusion matrix $Q_{ij}$ with
zero eigenvalue.
If the function ${\cal F}(x)$ is a constant, then
$\tilde D_L(\{a\})=\tilde D_e(\{a\})$ is the equilibrium distribution, otherwise it is a nonequilibrium
stationary distribution.

One instructive case for the nonequilibrium stationary solutions is that of $g=0$.
For this case, $Q_{ij}$ becomes proportional to the dynamic matrix $\Omega_{ij}$: 
$Q_{ij}=(T/\gamma) K_{i\al}K_{j\al}=(T/\gamma \omega^2)\Omega_{ij}$.
By the same argument as above $\Omega_{ij}$ is singular as well.
Note from (\ref{EQ36}) that $C^L_a(z)=(T/\omega^2)\cdot (1-\delta^*)/z$ in the limit of 
$z \rightarrow 0$. The other correlators  do not diverge at $z=0$.
Hence the model is nonergodic for $0 \leq \delta^* < 1$: the system is always 
driven into the nonergodic state in the linear case ($g=0$).
In this case the thermal noise alone is not enough to drive the system
to the equilibrium state.  This case is somewhat reminiscent of the
ideal gas case or the collection of independent harmonic oscillators
where the systems are trivially non-ergodic due to the absence of  interactions. 
Only when the nonlinear reversible mode coupling is present, as $T$ increases, the thermal noise can 
drive the system to the equilbrium state, hence making the system ergodic.
The onset temperature at which the ergodicity is recovered is the dynamic transition temperature.

In any event, further numerical and theoretical studies of possibile
ergodic-to-nonergodic transitions for nontrivial case $g \neq 0$ are warranted.

\section{Summary}

We have constructed a dynamic mean-field-type model involving $N$-component density and 
$M$-component velocity variables with reversible mode coupling and trivial Hamiltonian.
The model is exactly solvable in the limit of $N,M \rightarrow \infty$ with keeping
the ratio $\delta^* \equiv M/N$ finite.
The model exhibits a sharp dynamic transition to a nonergodic state only in the range
$0 \leq \delta^* <1$. 
The nature of the nonergodic state can be understood in terms of the nonequilibrium stationary 
solution of the Fokker-Planck equation for the probability distribution for the density variable.
It would be interesting to investigate the nonequilibrium aging behavior of the model.

\ack
We thank W. G\"{o}tze, J. J\"{a}ckle,  A. Latz, S. J. Lee,  and R. Schilling for useful 
suggestions and discussions.
BK is supported by the Interdisciplinary Research Program of the KOSEF (Grant No.
1999-2-114-007-3). 
The research of KK is supported by the Department of Energy, under contract
W-7405-ENG-36. An additional partial support to KK by the Cooperative
Research under the Japan-U.S. Cooperative Science Program sponsored by Japan Society
of Promotion of Science is  also gratefully acknowledged.

\bigskip

\end{document}